\documentclass[%
 reprint,
superscriptaddress,
 amsmath,amssymb,
 aps,
prapplied, floatfix, longbibliography ]{revtex4-2}

\usepackage[version=3]{mhchem} 
\usepackage{graphicx}
\usepackage{dcolumn}
\usepackage{bm}
\usepackage[colorlinks,linkcolor=blue,anchorcolor=blue,citecolor=blue,urlcolor=blue]{hyperref}
\usepackage[mathlines]{lineno}
\usepackage{amssymb}
\usepackage{amsmath}
\usepackage{natbib}
\usepackage{soul,color}
\usepackage{xcolor,colortbl}
\usepackage{multirow}
\usepackage{color}%

\newcommand{\modR}[1]{\textcolor{red}  {#1}}

\begin{document}

\title{Exotic properties and manipulation in 2D semimetal $\alpha$-Mn${_2}$B${_2}$(OH)${_2}$: a theoretical study}

\author{Pingwei~Liu}
\affiliation{School of Physics,
             Southeast University,
             Nanjing, 211189, PRC}

\author{Dan~Liu}
\email      {liudan2@seu.edu.cn}%
\affiliation{School of Physics,
             Southeast University,
             Nanjing, 211189, PRC}

\author{Shixin~Song}
\affiliation{School of Physics,
             Southeast University,
             Nanjing, 211189, PRC}

\author{Kang~Li}
\affiliation{School of Physics,
             Southeast University,
             Nanjing, 211189, PRC}

\author{Xueyong~Yuan}
\affiliation{School of Physics,
             Southeast University,
             Nanjing, 211189, PRC}%
\affiliation{Key Laboratory of Quantum Materials and Devices of Ministry of Education,
             Southeast University,
             Nanjing, 211189, PRC}

\author{Jie~Guan}
\email      {guanjie@seu.edu.cn}%
\affiliation{School of Physics,
             Southeast University,
             Nanjing, 211189, PRC}
			
\date{\today}

\begin{abstract}

Most functional materials possess one single outstanding property
and are limited to be used for a particular purpose. Instead of
integrating materials with different functions into one module,
designing materials with controllable multi-functions is more
promising for the electronic industry. In this study, we
investigate an unexplored $\alpha$-phase of two-dimensional (2D)
Mn$_{2}$B$_{2}$(OH)$_{2}$ theoretically. Eighteen distinct
electrical polarizations, characterized by three different
magnitudes and twelve different directions, are found in this
phase. The switch of the electrical polarizations is also linked to
an observed splitting of band structures between different spin
states and the ferroelasticity of the system. The manipulation of
these properties can be realized through controlling the alignment
of Mn-OH-Mn chains. Additionally, the approximately honeycomb
lattice for the atomic layer of boron indicate the potential
superconductivity in the system. The diverse and tunable properties
make the proposed material as an outstanding candidate for sensing
applications at the 2D limit.

\end{abstract}
\maketitle

\section{Introduction}

Since the successful isolation of graphene in 2004~\cite{FA2004},
2D materials have garnered considerable attention and efforts in
both fundamental research and industry applications. Over the past
decade, a myriad of 2D functional materials has been synthesised in
experiment or predicted in theory, such as 2D ferroelectric
materials~\cite{{BS21},{ding2017},{Kal15},{Fei18},{DT291},{cui18},{duan20},{zeng21}},
ferromagnetic
materials~\cite{{xu18},{pab18},{xu17},{novo19},{bota21},{he22}},
superconductors~\cite{{dean19},{ren15},{pas16},{iwa17},{xiong21}},
etc. To broaden the application of 2D functional materials, efforts
have being invested into integrating 2D materials with different
functionalities within single system, such as
heterostructures~\cite{{neto16},{duan22},{novo19},{ago17},{gogo17},{lz19}}.
Further exciting advances in this field include the realization of
magnetoelectric multiferroics through the coupling between the
magnetic polarization and the spin helix~\cite{{Song22},{Chao24}}
and the creation of intrinsic electric dipoles in 2D magnetic
bilayers by a relative rotation~\cite{{Xiang23},{BS23}}. This
direct method has achieved significant progress in artificial
regulation of physical properties. However, the strict preparation
conditions and the relatively weak coupling between the composed 2D
materials make this method difficult to apply on a large scale.
Thus, searching 2D materials with intrinsic and controllable
multiple properties remains an exciting topic of research.

\begin{figure*}[t]
\includegraphics[width=1.7\columnwidth]{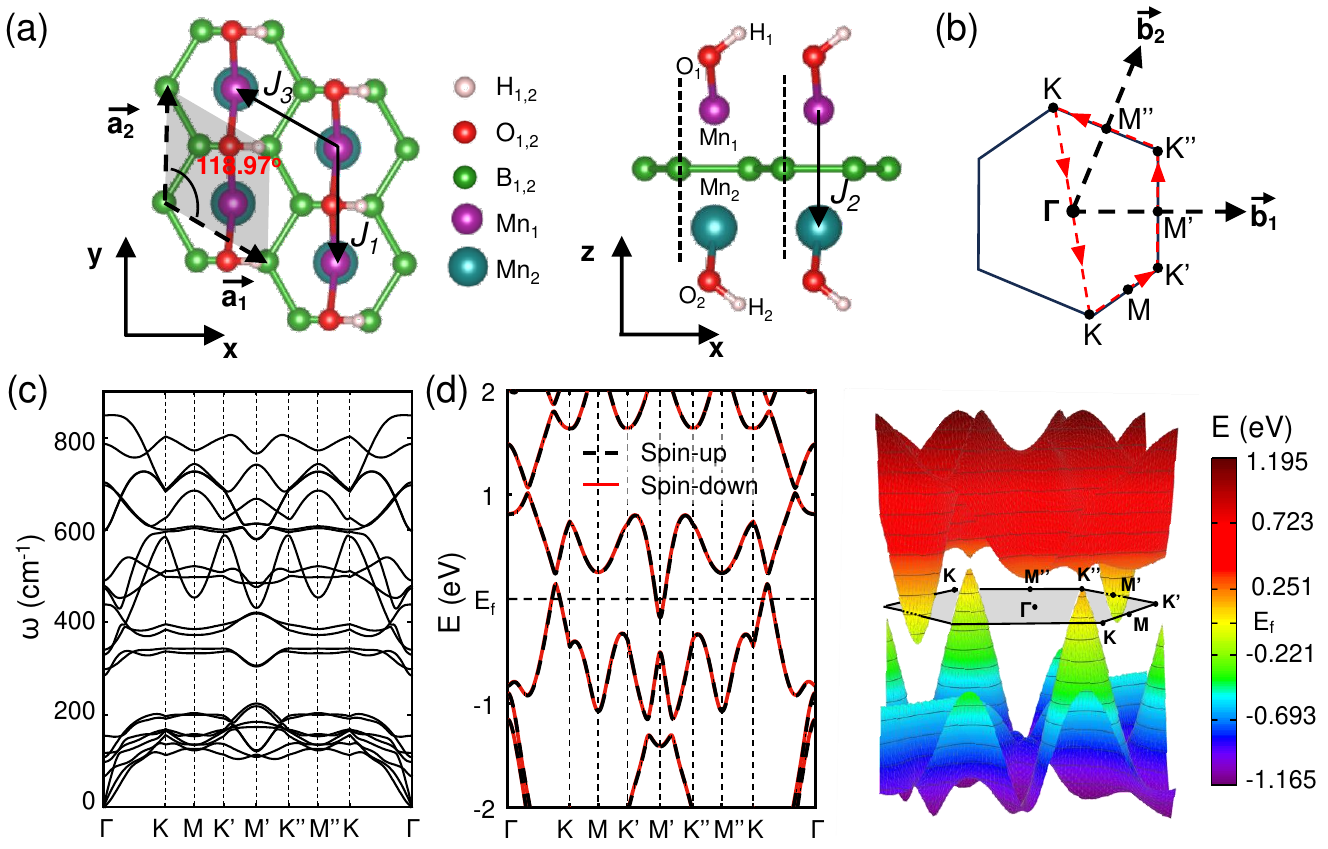}
\caption{%
(a) Atomic structure of 2D Mn$_{2}$B$_{2}$(OH)$_{2}$. Mn atoms with
different spins are presented in purple and green. The unit cell is
highlighted by the transparent grey area which is a slightly
distorted rhombus. The corresponding Brillouin zone and high
symmetry points are exhibited in (b). DFT-PBE calculated phonon
spectra is presented in (c). (d) Spin polarized electronic band
structure and three-dimensional plot of the two bands below and above the fermi level.
\label{fig1}}
\end{figure*}


Here, {\em ab initio} calculations unveil diverse tunable ferroic
properties in a previously unexplored $\alpha$-phase of 2D material
Mn$_{2}$B$_{2}$(OH)$_{2}$ $(MBOH)$. This material displays a
ferroelectric $(FE)$ behavior originating from the ordered
arrangement of hydroxyl radicals bridging the neighboring Mn atoms.
The formed Mn-OH-Mn ($MOM$) chains distribute on both sides of a
flat monolayer of boron with a nearly-honeycomb lattice $(nh-B)$.
The electrical polarization of each individual side can have six
different orientations and the system exhibits eighteen distinct
polarization states in total, which are manipulable by altering the
direction of $MOM$ chains. Within this polarization manipulating
process, this 2D material shows a ferroelastic behavior, indicating
a coupling between ferroelectricity and ferroelasticity. The
$\alpha$-phase $MBOH$ exhibits a preference of antiferromagnetism,
with magnetic moments of Mn atoms on different sides pointing in
opposite directions. By independently controlling the $MOM$ chains
on each side of $nh-B$, the degeneracy of the electronic states in
spin-up and spin-down channels can be regulated.

\section{Computational Techniques}

Our calculations of the atomic structures, electrical and magnetic
properties have been performed using DFT as implemented in the
{\textsc{VASP}}~\cite{VASP,VASPPAW} code. We used the
Perdew-Burke-Ernzerhof (PBE)~\cite{PBE} exchange-correlation
functional. The DFT+U approximation is employed with U=3~eV and
J=1~eV imposed on $3d$ orbitals of Mn~\cite{GGAU}. Periodic
boundary conditions have been used throughout the study, with
monolayers represented by a periodic array of slabs separated by a
20~{\AA} thick vacuum region. The calculations were performed using
the projector augmented wave (PAW) method~\cite{VASPPAW} and
$560$~eV as energy cutoff. The reciprocal space has been sampled by
a fine grid~\cite{Monkhorst-Pack76} of
$12{\times}12{\times}1$~$k$-points in the 2D Brillouin zone (BZ).
All geometries have been optimized using the conjugate gradient
(CG) method~\cite{CGmethod}, until none of the residual
Hellmann-Feynman forces exceeded $10^{-2}$~eV/{\AA}. The phonon
spectrum was calculated using a $6{\times}6{\times}1$ supercell,
and the real-space force constants of supercells were calculated
using density-functional perturbation theory (DFPT) as implemented
in VASP~\cite{togo2015}.

\section{Results}

\subsection{Atomic structure, stability and electronic properties}

Commencing with the precursor material MnB~\cite{han23}, our
investigation explores various possible allotropes of $MBOH$ with
different magnetic orders and emphasize on the most stable one,
named as $\alpha$-phase, are shown in Fig.~\ref{fig1}(a).
\modR{The strategy and details used to exploring the $-$OH
decorated MnB is presented in the Appendix A.} In $\alpha$-phase
$MBOH$, an atomic monolayer of boron with a nearly-honeycomb
lattice is established. The adjacent Mn atoms, interconnected via
hydroxyl radicals ($-$OH), are distributed on both sides of the
boron layer. The unit cell contains two boron atoms, labelled as
$B_{1,2}$, two Mn atoms, and two $-$OH radicals. The subscripts $1$
or $2$ for Mn, O, H means whether an atom lies above (1) or below
(2) the $nh-B$ as shown in the right panel of Fig.~\ref{fig1}(a).
The lattice of the $\alpha$-phase is a slightly distorted hexagonal
lattice, with lattice constants $a_{1}=3.14~\AA$, $a_{2}=3.04\AA$,
and an angel of $118.97^{\circ}$ between $\vec{a_1}$ and
$\vec{a_2}$. Consequently, the first $BZ$ is also a distorted
hexagonal lattice, as shown in Fig.~\ref{fig1}(b) with the
distortion exaggerated. To validate the stability of the
$\alpha$-phase, we calculated the phonon spectrum and the results
are shown in Fig.~\ref{fig1}(c). No imaginary frequency is
detected, which means $\alpha$-phase is dynamically stable.
\modR{We have also confirmed the dynamic stability of the
$\alpha$-phase by performing canonical molecular dynamics (MD)
simulation at room temperature of 300~K of 5~ps which are
presented in Appendix C.}

DFT-PBE calculations of electronic band structure are presented in
Fig.~\ref{fig1}(d). The resutls indicate that the $\alpha$-phase
$MBOH$ is a semimetal, in which electron and hole pockets coexist
on the Fermi surface. Analyzing the 2D band structure in
Fig.~\ref{fig1}(d), we observe that the valence band maxima (VBM)
resides at the pathway from $\Gamma$ to $K$, while conduction band
minimum (CBM) is located at the $M'$. To ensure no crucial
electronic information is overlooked in the 2D band structure, a
three-dimensional band structure was also calculated to provide a
comprehensive view of all details. The dispersion of the two bands
above and below the Fermi surface across the entire $k$-space is
exhibited in the right panel of Fig.~\ref{fig1}(d). \modR{In the
most stable $\alpha$-phase, an interesting antiferromagnetic (AFM)
order called A-type AFM (A-AFM) is found with antiferromagnetically
coupled ferromagnetic Mn layers.} The band structure exhibits a
typical anti-ferromagnetic ($AFM$) behavior, where band dispersion
for spin-up and spin-down states are degenerate. We also analysis
the magnetic information by calculating the spin-polarized
projected density of state, as presented in Fig.~\ref{fig9} of
Appendix B. \modR{It confirms that the magnetic moments are mainly
contributed by the layer of Mn$_1$ and Mn$_2$ which point in
opposite directions.}

\begin{figure}[t]
\includegraphics[width=1.0\columnwidth]{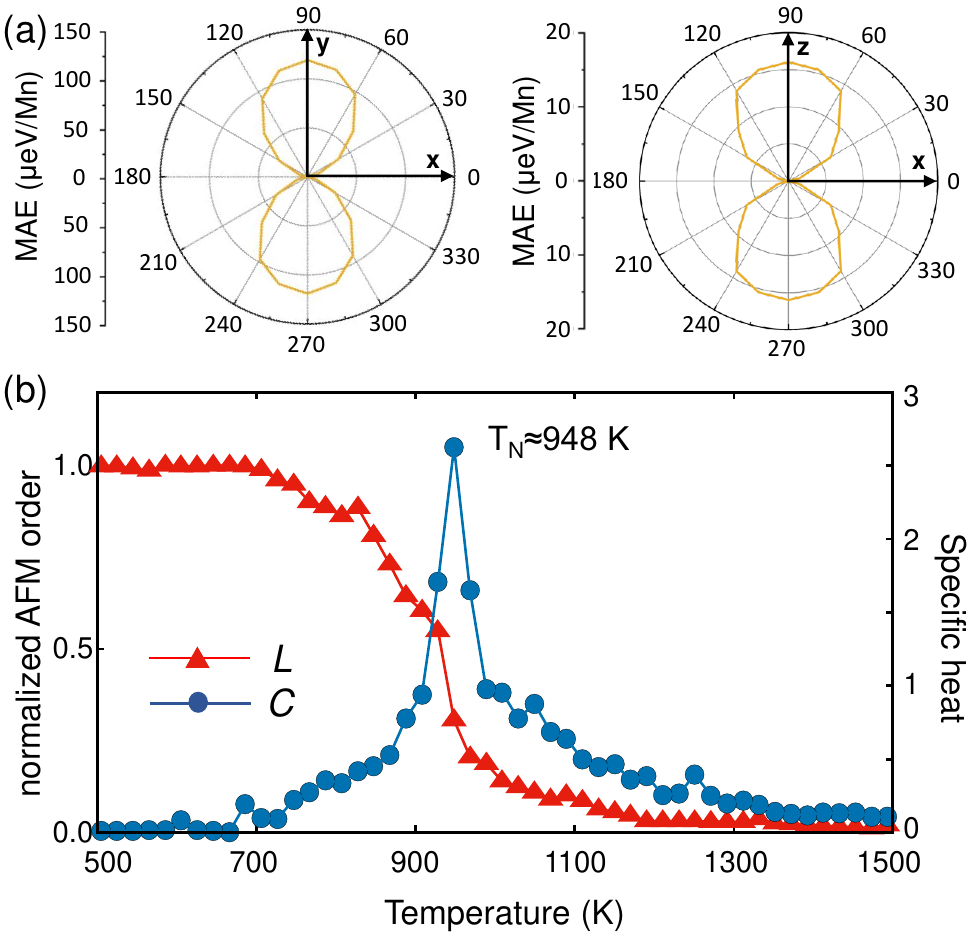}
\caption{%
\modR{(a) The DFT calculated MAE as a function of spin orientation. 
(b) MC simulated specific heat and normalized AFM order %
parameter L as a function of temperature. The AFM order %
parameter is defined as L=S$_u$-S$_l$, where S is the normalized spin %
and u/l denotes the upper and lower layers of Mn bilayers.}%
\label{fig2}}
\end{figure}

\modR{The magnetic anisotropy is also calculated by rotating the spin
orientation. As shown in Fig.~\ref{fig2}(a), the magnetic easy axis lies on the
atomic plane and is along the x-axis. The magnetocrystalline
anisotropy energy (MAE) is estimated to be around 120~$\mu$eV/Mn atom.
As shown in Fig.~\ref{fig1}(a), here we consider the
nearest-neighbor exchange constant J$_1$, next-nearest-neighbor
exchange constant J$_2$, and next-next-nearest-neighbor exchange
constant J$_3$ to determine the magnetic order of the system. Based
on the optimized structure of its ground state, through mapping the
DFT energy to the Heisenberg model with normalized spins $|S|=1$,
these magnetic exchange coupling constants are derived to be
12.79~meV, -55.47~meV and 2.52~meV. Both J$_1$ and J$_3$ indicating
a intralayer ferromagnetism, while J$_2$ prefers a interlayer
antiferromagnetim of Mn$_1$ and Mn$_2$ atomic layers. Such a
configuration of J is not frustrated, which co-stabilize the
layered A-AFM order. Based on the above exchange coupling
coefficients, the MC method was employed to simulate the magnetic
transition. Its Neel temperature T$_N$ is estimated to be 948 K, as
indicated by the peak of specific heat shown in Fig.~\ref{fig2}(b).}

\begin{figure}[t]
\includegraphics[width=1.0\columnwidth]{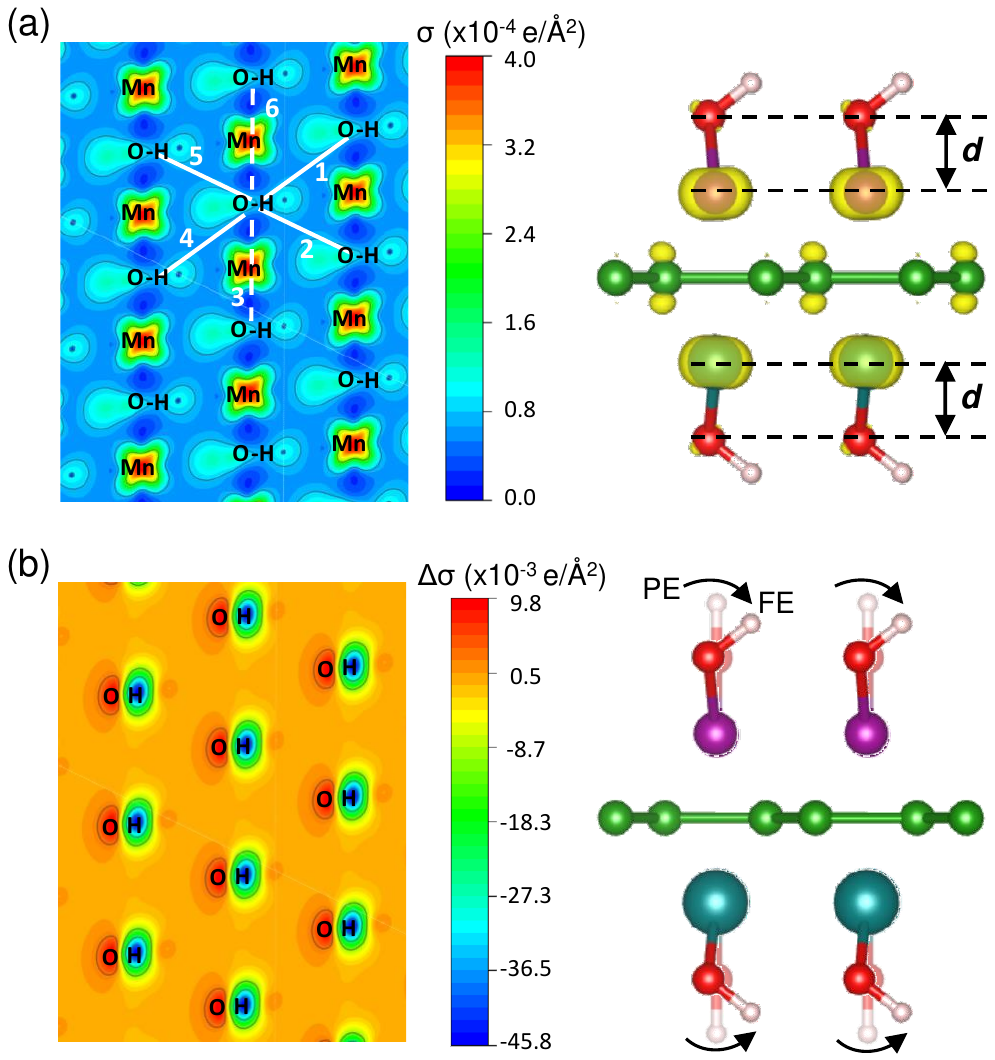}
\caption{%
\modR{(a) Partial electron densities contour maps for $\alpha$-phase MBOH %
with polarization state of $\vec{UP_1}$+$\vec{LP_1}$ taken through %
atomic plane. %
The electronic spacial distribution at the fermi level is shown in the right panel. %
(b) Charge density difference of between FE state %
$\vec{UP_1}$+$\vec{LP_1}$ and paraelectric (PE) state of $\alpha$-phase MBOH. %
The transformation from the PE state to the FE state is illustrated in the right panel.} %
\label{fig3}}
\end{figure}

\subsection{Ferroic properties and tunable electrical polarization}

As seen in the atomic structure, the $-$OH radicals are oriented in
the direction perpendicular to the $MOM$ chains and partially
tilted toward to $nh-B$. Due to the orderly arrangement of dipole
moments of the -OH radicals, a macroscopic polarization is
established. As the -OH radicals are distributed on both sides of
the $nh-B$, the net polarization comprises polarizations from both
the upper $-$OH radicals, named as upper polarization ($\vec{UP}$),
and the lower ones, named as lower polarization ($\vec{LP}$). Here,
we employ a point charge model to evaluate the value of the
$\vec{UP}$ and $\vec{LP}$. We do bader charge analysis to get the
net charge $\Delta{Q}$ of cations and anions, and measure the
distance $d$ between the center of positive charge and negative
charge. The values of $\vec{UP}$ and $\vec{LP}$ are then evaluated
as $P={\Delta}{Q}\times{d}=1.03~pC/cm$.

\modR{In spite of using point charge model to evaluate the polarization,
it is necessary to prove the existence of the polarization, since
the electrostatic forces may be strongly screened by the itinerant
electrons in the metallic $\alpha$-phase MBOH. In
Fig.~\ref{fig3}(a), we sketched the electron densities of
$\alpha$-phase MBOH around the Fermi level on the atomic planner to
have direct view of the charge distribution. We found the
conduction charge is mainly spacing around Mn atoms, and rarely
spacing around O and H atoms, so we can conclude that the $-$OH is
a bare ionic radical which is confirmed by the spatial electron
distribution exhibited in the right panel of Fig.~\ref{fig3}(a). As
mentioned above, the in-plan electrical polarization is coming from
the tilting $-$OH radicals. From Fig.~\ref{fig3}(a), since there is
little conduction charges distributed between pair-1, 2, 4 and 5,
the screening effect of dipole-dipole interaction in these pairs is
small. We should be careful about the screening effect of the
pair-3 and 6. It seems that the dipole interactions these two
dipole pairs will be screened by the itinerant electrons around Mn
atoms. However, the $-$OH dipoles and the Mn atoms are separated
vertically with distance $d\sim$1.5~$\AA$, we expect the screening
effect actually is limited. This is confirmed by the calculation of
the charge difference induced through tilting $-$OH radicals from
the PE state to the FE state, which is illustrated in the right
panel of Fig.~\ref{fig3}(b). As shown in left panel of
Fig.~\ref{fig3}(b), the charge density change $\Delta\sigma$ is
plotted. As seen from this plot, we actually did not see a
significant redistribution of itinerant electrons around Mn atoms,
which affirms conversely the dipole interaction in pair-3 and 6 is
slightly screened by the conduction electrons of Mn atoms.}

\begin{figure}[t]
\includegraphics[width=1.0\columnwidth]{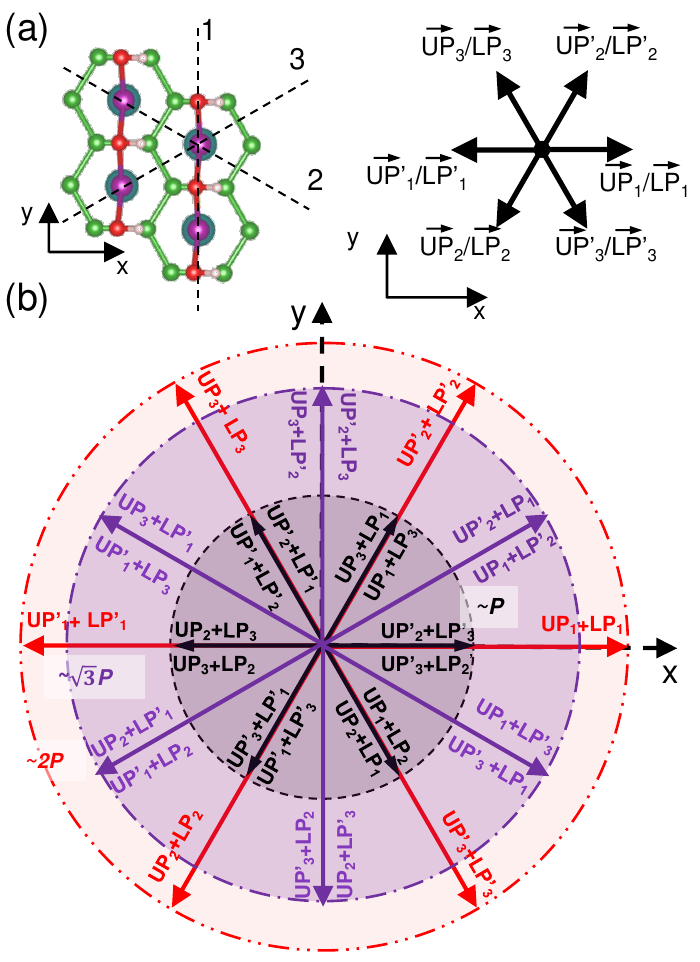}
\caption{%
(a) Three directions along which $MOM$ chains could line up, as presented by black dashed lines (left panel), and all possible
directions of upper polarization ($\vec{UP}$) and lower polarization
($\vec{LP}$).
(b) Eighteen non-zero electrical polarizations of the system,
polarization with magnitude of P, $\sqrt{3}P$ and 2P are indicated by
black solid arrow, purple solid arrow and red solid arrow.
\label{fig4}}
\end{figure}

In Fig.~\ref{fig4}(a), the $MOM$ chains lying on both sides of the
$nh-B$ layer align along the same direction, which is marked as
direction `1' and is same direction as $\vec{a_{2}}$ shown in
Fig.~\ref{fig1}(a). In the view of stability and physical
properties, there are two other equivalent directions, marked as
`2' and `3', for $MOM$ chains to line up along. For each direction,
the $-$OH radicals can orient to the right side or the left side of
$MOM$ chains. We denote the polarization states as
$\vec{UP_n}/\vec{LP_n}$ and $\vec{UP'_n}/\vec{LP'_n}$, where n=1,
2, 3, indicating the orientation of the $MOM$ chains and with or
without $'$ means the $-$OH radicals orient to the right or left
side of $MOM$ chains. Totally, there are six polarization states
for $\vec{UP}$ and $\vec{LP}$, as shown in the right panel of
Fig.~\ref{fig4}(a). The net polarization of the system results from
the combination of $\vec{UP}$ and $\vec{LP}$
\modR{($\vec{UP^{(')}_n}+\vec{LP^{(')}_m}$, n, m=1, 2, 3),} leading to
thirty-six possible combinations. The combined polarization has
four possible values of 0, $P$, $\sqrt{3}P$, $2P$ and each non-zero
value has six possible corresponding directions.
\modR{So, there are totally eighteen possible non-zero net polarizations
of the system as shown in Fig.~\ref{fig4}(b). Each net polarization
with magnitude of $P$, $\sqrt{3}P$ has two possible combinations,
while each net polarization with magnitude of $2P$ has one possible
combinations. Instead of the term of "possible combination", in the
following we will use the term of "polarization states". As a
result, in $\alpha$-phase MBOH, there are thirty polarization
states possess non-zero net polarization.} Based these results, one
interesting question is arisen that could the $\vec{UP}/\vec{LP}$
be switched between different states? Since the polarization state
is determined by the alignment of $MOM$ chains and the orientation
of $-$OH radicals, we thus categorize the transformation of
polarization states into two parts, one is changing the alignment
of $MOM$ chains, the other is flipping the orientation of $-$OH
radicals.

\begin{figure*}[t]
\includegraphics[width=1.7\columnwidth]{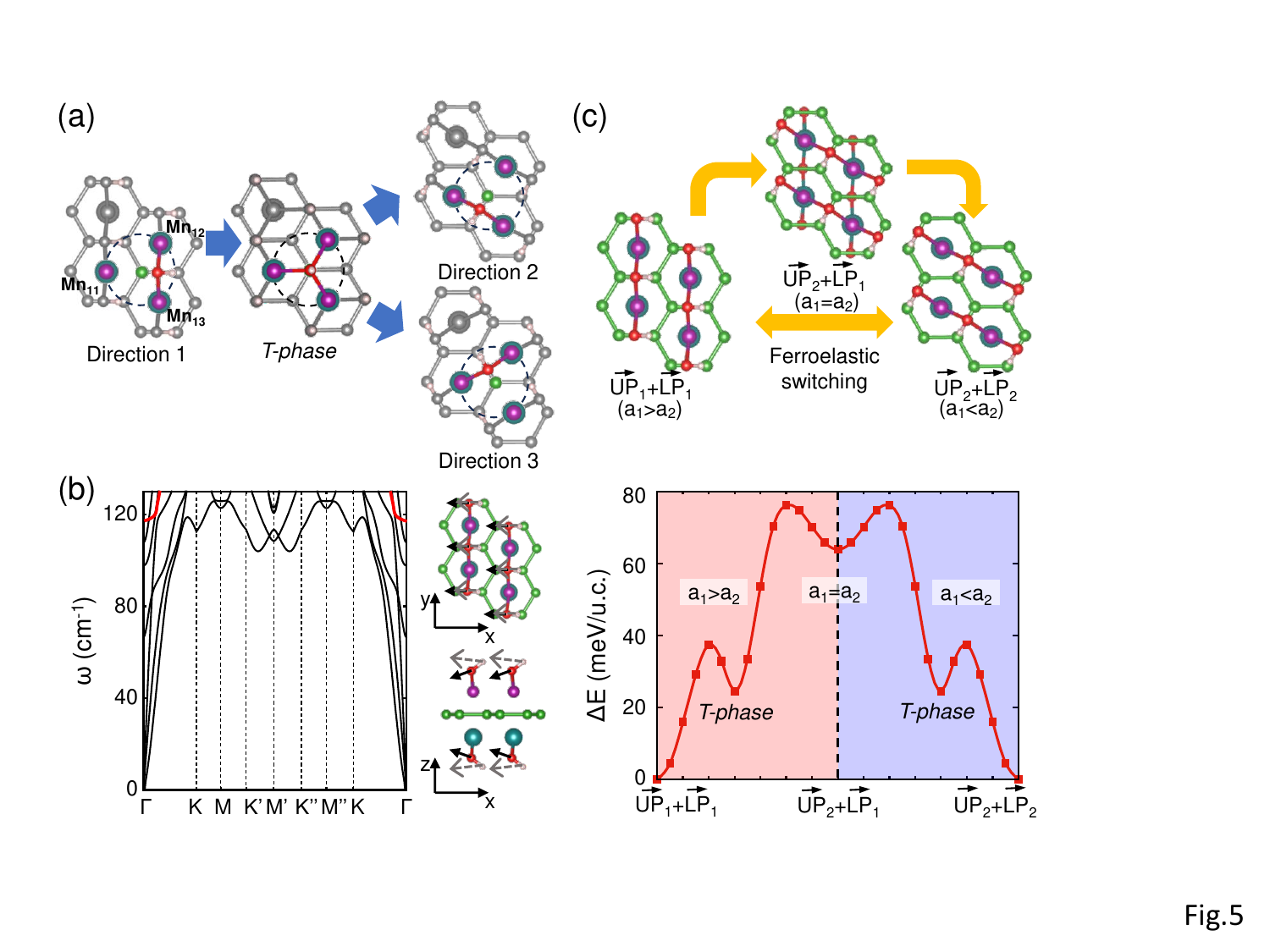}
\caption{%
(a) Schematic mechanism of changing the alignment of the upper
$MOM$ chains from direction `1' to direction `2' or `3' while
fixing the lower $MOM$ chains, which undergos a transition state
named as `T-phase'. (b) Vibration mode of the special branch of
phonon spectra, which is highlighted by red color,
indicating the
movements of $-$OH radicals toward the top of the neighboring B atoms.
(c) Illustration of changing polarization state of the system from
$\vec{UP_1}+\vec{LP_1}$ to $\vec{UP_2}+\vec{LP_2}$ and the
corresponding DFT-PBE energy difference $\Delta{E}$ encountered
during the switching process.%
\label{fig5}}
\end{figure*}

\begin{figure*}[t]
\includegraphics[width=1.7\columnwidth]{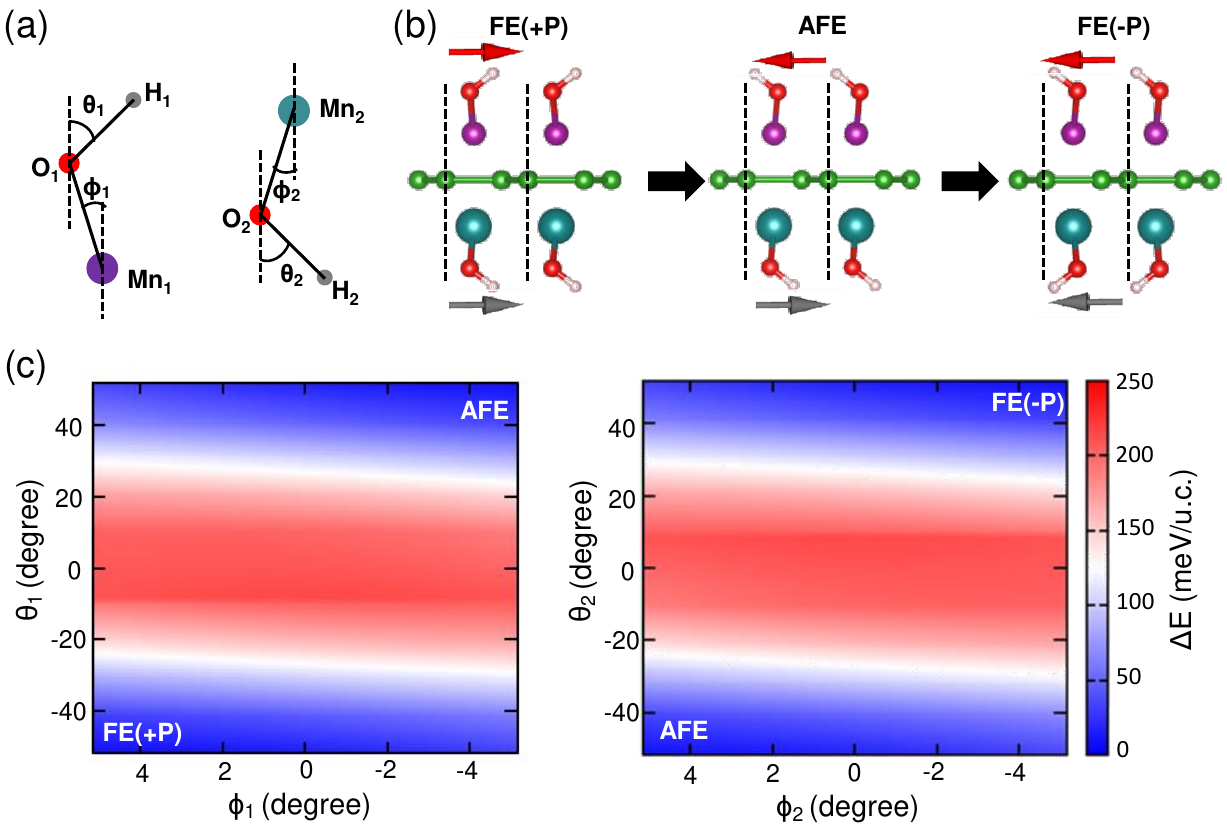}
\caption{%
(a) Dihedral angle $\phi$ and angle $\theta$ used to characterize
the position of $-$OH radicals. (b)
Structural change during the polarization flipping path starting
from the initial state with polarization $+P$, across the $AFE$ state,
and reaching the finial state with polarization of $-P$. The
polarization directions are indicated by red and grey arrows.
(c) Corresponding energy changes $\Delta{E}$ as a
function of $\phi$ and $\theta$ during the process of polarization flipping. Subscripts
$1$ and $2$ represent the $-$OH radicals in the upper layer and
lower layer of boron atomic layer, respectively.%
\label{fig6}}
\end{figure*}

\begin{figure}[t]
\includegraphics[width=1.0\columnwidth]{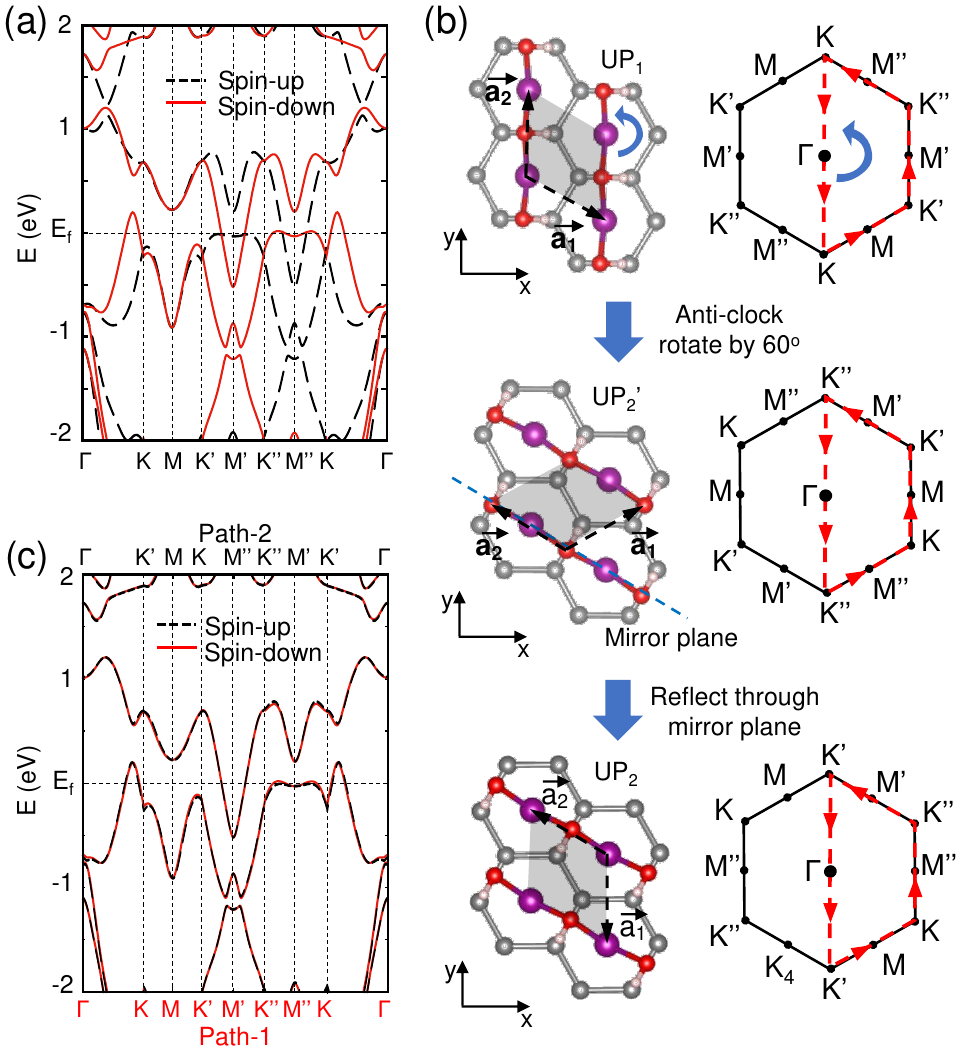}
\caption{%
(a) Spin polarized electronic band structure of $\alpha$-phase
$MBOH$ with polarization state of $\vec{UP_2}+\vec{LP_1}$. (b)
Atomic transformation of the system while the $\vec{UP}$ changing
from $\vec{UP_1}$ to $\vec{UP_2}$, and accompanied with the
transformation of primitive unit-cell and it's first Brillouin
zone. (c) Reconstruction of band structure with spin-up state along
the k-path of $\Gamma$-$K'$-$M$-$K$-$M''$-$K''$-$M'$-$K'$-$\Gamma$
and comparing with band structure with spin-down state along the
k-path of $\Gamma$-$K$-$M$-$K'$-$M'$-$K''$-$M''$-$K$-$\Gamma$.%
\label{fig7}}
\end{figure}

For illustrating how to change the alignment of $MOM$ chains, here
we propose a plausible scenario in Fig.~\ref{fig5}(a). We suppose
that a $MOM$ chain on the upper side of the $nh-B$ is along
direction `1' initially. The $-$OH radical connecting Mn$_{12}$ and
Mn$_{13}$ firstly swings to the top of B atom, and then move
further to connect Mn$_{11}$ and Mn$_{13}$, forming a $MOM$ chain
along direction `2', to connect Mn$_{11}$ and Mn$_{12}$, forming a
$MOM$ chain along direction `3'. The validity of this scenario
depends on the initial movement of $-$OH towards the top of B,
which is supported by the vibration mode of the seventh branch as
highlighted by red color in Fig.~\ref{fig5}(b). This mode is
characterized by the swing of the $-$OH radical towards the top of
a boron atom. The frequency of this vibration mode at the
$\Gamma$-point is $117.33~cm^{-1}$, which is about $0.56$ times of
room temperature, meaning that the cost of an external field used
to trigger this scenario is acceptable.

Within this model of switching $\vec{UP}/\vec{LP}$, we investigate
the energy change $\Delta{E}$ during the process of transforming
from $\vec{UP_1}$+$\vec{LP_1}$ to $\vec{UP_2}$+$\vec{LP_2}$, as
shown in Fig.~\ref{fig5}(c). In the initial
$\vec{UP_1}$+$\vec{LP_1}$ state, all $MOM$ chains are oriented
along direction `1', with the lattice constant $a_{2}<a_{1}$. The
$\vec{UP_2}$+$\vec{LP_1}$ state is then obtained by twisting the
upper $MOM$ chains to direction `2', with the lattice constant
$a_{2}=a_{1}$. Finally, the lower $MOM$ chains change to direction
`2', reaching the $\vec{UP_2}$+$\vec{LP_2}$ state, with the lattice
constant $a_{2}>a_{1}$. The highest energy barrier is about
75~meV/u.c.. The state of $\vec{UP_1}$+$\vec{LP_1}$ is
energetically equally stable as $\vec{UP_2}$+$\vec{LP_2}$, so this
polarization switching process is also a ferroelastic process. In
other words, the ferroelectricity is coupled with the
ferroelasticity and the system exhibts a multiferroic behavior.
Besides the stability of the $\alpha$-phase with upper and lower
$MOM$ chains aligning along the same direction shown in
Fig.~\ref{fig1}(c), we also present the stability of the system
with upper and lower $MOM$ chains aligning along different
directions in Fig.~\ref{fig10} of Appendix C.

To flip the orientation of the $-$OH radicals with the bond length
and bond angle fixed, we vary the dihedral angles
$\phi_{1}$/$\phi_{2}$ between Mn-O-Mn plane and y-z plane, and the
angle $\theta_{1}$/$\theta_{2}$ between the O-H bond and the
z-direction as defined in the Fig.~\ref{fig6}(a). Taking the
transformation from $\vec{UP_1}$+$\vec{LP_1}$ to
$\vec{UP'_1}$+$\vec{LP'_1}$ as an example, we study the energy cost
during the process in which the net polarization is reversed. This
process is divided into two steps. The first step is changing
$\vec{UP_1}$ to $\vec{UP'_1}$ through rolling over the
$-$O$_1$H$_1$ radicals at the upper side. The system starts from
the initial FE state (FE(+P)) and arrives at an anti-ferroelectric
(AFE) state. The second step is changing $\vec{LP_1}$ to
$\vec{LP'_1}$ through rolling over the $-$O$_2$H$_2$ radicals at
the lower side. The system arrives at an FE state with opposite
polarization (FE(-P)). The process is illustrated in
Fig.~\ref{fig6}(b). We calculated the energy change $\Delta{E}$ of
the system as a function of $\theta$ and $\phi$ for the two steps
of polarization flipping. From the energy map presented in
Fig.~\ref{fig6}(c), we observe that most of energy cost occurs
while changing $\theta$, and the highest energy cost is about
$\Delta{E}=225~meV/u.c.$.

\subsection{Spin polarization accompanied with change of polarization state}

As shown in Fig.~\ref{fig1}(d), the electronic band structure
associate with spin-up and spin-down states are degenerate for the
system with polarization state of $\vec{UP_1}$+$\vec{LP_1}$.
However, the electronic state can be tuned by switching
polarization states of the system. An illustrative example is
presented in Fig.~\ref{fig7}(a) for the system featuring the
polarization state of $\vec{UP_2}$+$\vec{LP_1}$. In this case, the
energetic degeneracy of spin-up and spin-down is lifted. More
interestingly, the dispersion relation of one spin state appears to
replicate the other one, but through a different k-path.

As mentioned above, the difference in the atomic structure between
the system with $\vec{UP_1}$+$\vec{LP_1}$ and
$\vec{UP_2}$+$\vec{LP_1}$ is mainly about the alignment of upper
$MOM$ chains, as depicted in Fig.~\ref{fig5}. Here we present a
two-step transformation process in geometry between the two
polarization states. As illustrated in Fig.~\ref{fig7}(b), we
initially rotate the upper $MOM$ chains counterclockwise by
$60^{\circ}$ to align them with direction `2', resulting in the
polarization of $\vec{UP'_2}$. Then, a reflection operation is
applied to the upper $MOM$ chains through a mirror plane marked by
the blue dashed line in Fig.~\ref{fig7}(b), leading to a change in
the upper polarization to $\vec{UP_2}$. During the rotating and
reflecting operation of upper $MOM$ chains, the atomic sub-unitcell
and its corresponding first $BZ$ of the upper $MOM$ chains are
transformed in the same way with respect to the fixed cartesian
coordinate.

The spin-down and spin-up electronic states originate from the
upper and lower $MOM$ chains, respectively, and they become
degenerate when the system is in the state of
$\vec{UP_1}$+$\vec{LP_1}$. In Fig.~\ref{fig7}(b), we manipulate the
first $BZ$ referring to the Cartesian coordinates of the upper
$MOM$ chains. If we go through
Path-2=$\Gamma$-$K'$-$M$-$K$-$M''$-$K''$-$M'$-$K'$-$\Gamma$ for the
$\vec{UP_2}$, the energetic dispersion relation of will be the same
as the energetic dispersion relation of $\vec{UP_1}$ through
Path-1=$\Gamma$-$K$-$M$-$K'$-$M'$-$K''$-$M''$-$K$-$\Gamma$. In
other words, since the electronic properties are degenerate for
$\vec{UP_1}$ and $\vec{LP_1}$, we can assert that the recombination
of the band structure of the spin-up states, originating mainly
from the lower $MOM$ chains, through $Path-2$ is a reproduction of
the energetic dispersion relation of spin-down states through
$Path-1$ for the system of $\vec{UP_2}$+$\vec{LP_1}$. To validate
this assertion, we compare the band structure of spin-down states
as a function of $Path-1$ and of spin-up states as a function of
$Path-2$ in Fig.~\ref{fig7}(d). The minimal difference observed
between them confirms the validity of our analysis. With this
clarification, it becomes evident that the electronic properties of
spin-up states at $K$($K'$), $M'$($M''$) are equivalent to those
with spin-down states at $K'$($K$), $M''$($M'$), respectively. We
present the electronic band structures with spin-up states and
spin-down states from $\Gamma$ point to each individual $k$ point
in Fig.~\ref{fig11} of Appendix D. We find that the dispersion
relations of spin-up states through $\Gamma-K$, $\Gamma-K'$,
$\Gamma-M'$ and $\Gamma-M''$ are identical to that of spin-down
states through $\Gamma-K'$, $\Gamma-K$, $\Gamma-M''$ and
$\Gamma-M'$.

Considering the splitting of the spin-polarized band structure is
decided by electronic spin and the electrons of H atoms are rarely
spin-polarized, we can deduce that either H atoms stay at the right
side or left side of the $MOM$ chains does not impact the band
structure splitting. We conducted band structure calculation of the
system with $\vec{UP'_2}$+$\vec{LP_1}$, in which H atoms stay at
the opposite side of the upper $MOM$ chains in comparison with
$\vec{UP_2}$+$\vec{LP_1}$. As shown in Fig.~\ref{fig12} of Appendix
E, the band structure splitting are the same as that for the system
with $\vec{UP_2}$+$\vec{LP_1}$.

\begin{figure}[t]
\includegraphics[width=1.0\columnwidth]{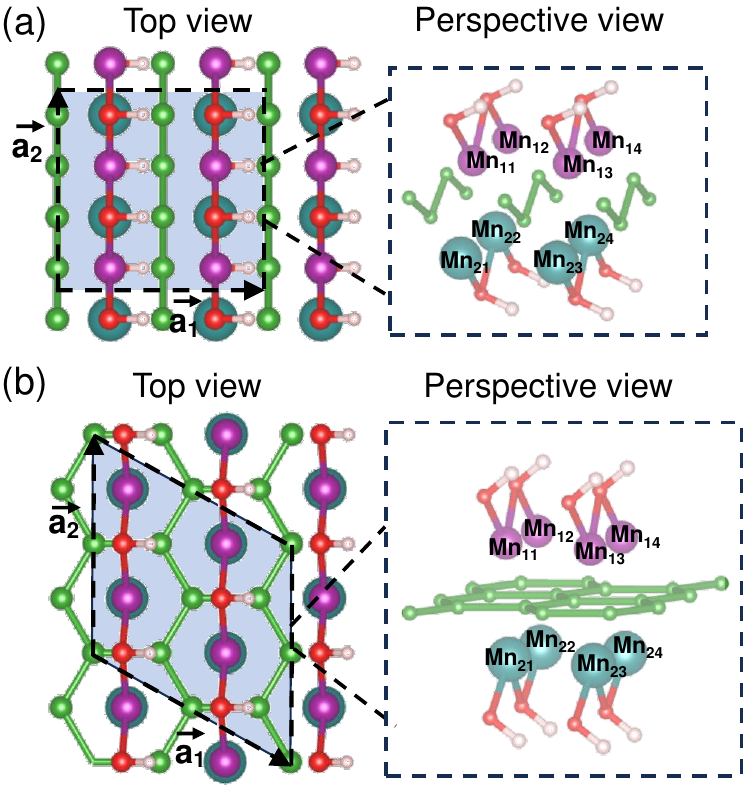}
\caption{ %
\modR{(a) Precursor structure MnB attached with $-$OH radicals used to %
searching stable MBOH with possible magnetic orders. %
(b) Precursor structure $\alpha$-phase MBOH with A-AFM magnetic %
order used to searching more stable MBOH with other possible magnetic orders. %
The unit cells used to do structure optimization are highlighted by the transparent blue areas, %
there are four upper Mn atoms (Mn$_{1n}$) and four lower Mn atoms (Mn$_{2n}$) (n=1, 2, 3, 4) contained in the unit cell.}%
\label{fig8}}
\end{figure}

\begin{figure}[t]
\includegraphics[width=1.0\columnwidth]{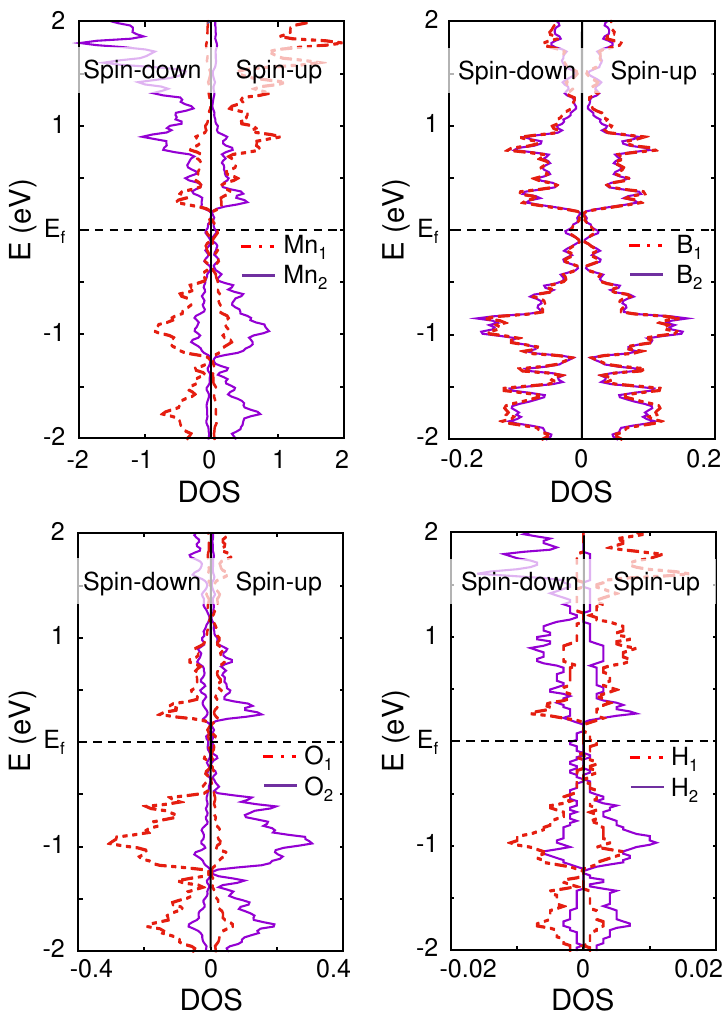}
\caption{ %
Spin-polarized projected density of state onto Mn$_1$, Mn$_2$, B$_1$, B$_2$, O$_1$, O$_2$, H$_1$, H$_2$. %
\label{fig9}}
\end{figure}

\begin{figure}[t]
\includegraphics[width=0.9\columnwidth]{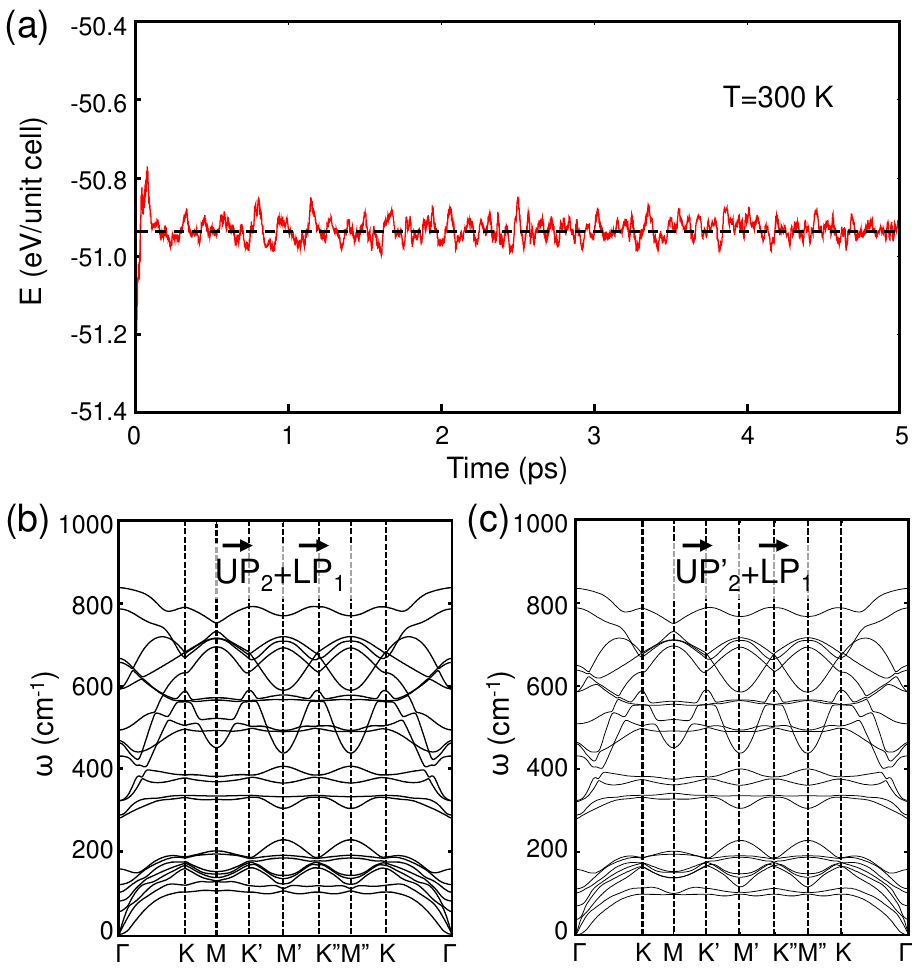}
\caption{ %
\modR{(a) Change in the potential energy E of polarization state of
$\vec{UP_2}+\vec{LP_1}$ during a 5-ps long canonical MD simulation
run at T=300~K.
(b) DFT-PBE calculated phonon spectra of polarization states of $\vec{UP_2}+\vec{LP_1}$ and $\vec{UP'_2}+\vec{LP_1}$.} %
\label{fig10}}
\end{figure}

\section{Discussion}

In addition to the physical properties explored above, another
noteworthy phenomenon is the formation of an atomic boron layer
with a nearly-honeycomb lattice. A presence of the honeycomb
lattice of boron is believed to the key factor for the
superconductivity in the boron system, such as
$MgB_{2}$~\cite{{mgb1},{mgb2}}. To construct the $MBOH$ structure,
we start from the initial structure of Mn$_2$AlB$_2$ and attach
$-$OH radicals on the MnB framework instead of the Al atoms. The
corresponding potential energy change during the conjugate gradient
(CG) optimization process is plotted in Fig.~\ref{fig13} of
Appendix F. A monotonic decrease of the potential energy is found
and almost no energy barrier is observed. This suggests that
obtaining the honeycomb lattice of boron is relatively
straightforward and no catalysts or other assisting methods are
needed in experiments. In comparison to synthesizing the honeycomb
lattice of boron through electron doping~\cite{DT272}, the use of
$-$OH radical attachment is more direct and applicable on large
scale.

We have observed the coexistence and coupling of diverse advanced
physical properties in 2D $\alpha$-phase $MBOH$. To manipulate
these properties effectively, it is crucial to control the $MOM$
chains and the -OH radicals in the way described in Fig.~\ref{fig6}
and Fig.~\ref{fig7}. A critical question arises, how can this
manipulation be achieved in experiments? Particularly, among these
eighteen polarization states, switching from one to the other
requests independent manipulation of -OH radicals of the upper or
lower layer, which are separated by 7~$\AA$. We propose here that
scanning tunneling microscopy may be a suitable technique for this
purpose~\cite{{STM1},{STM2}}. Through controlling the bias voltage,
STM can generate an electrical field to change the direction of
$MOM$ chains in the upper layer while having no effect on the $MOM$
chains staying in the lower layer. This approach provides a
potential experimental avenue for achieving the desired
manipulation.

While all the eighteen polarizations in this study lie in plane, we
observe a unique case of the T-phase, as illustrated in
Fig.~\ref{fig5}(c). This phase, which is about 24.6~meV/u.c. less
stable the $\alpha$-phase, features -OH radicals on one side
perpendicular to the plane and that on the other side partially lie
in the plane, resulting in a net out-of-plane polarization. This
out-of-plane polarization can be flipped by exchanging the
orientation modes of -OH radicals of both sides. When combined with
these out-of-plane polarizations, the polarization states of the 2D
$\alpha$-phase $MBOH$ becomes more attractive.

\modR{In the main text, we calculate the electronic band structure of the
polarization state of $\vec{UP_1}+\vec{LP_1}$ and
$\vec{UP_2}+\vec{LP_1}$ to study the effect of the alignment of
upper and lower MOM chains on the electronic properties, and we see
the spin splitting occurs while the upper and lower MOM chains
lying in different directions. In addition, we also calculate the
band structure of $\vec{UP'_2}+\vec{LP_1}$ to study the effect of
the tilting of $-$OH radical on the electronic properties, and we
see the effect is negligible. Besides these three polarization
states, there are other twenty-seven polarization states mentioned
in Fig.~\ref{fig4}(b), the electronic properties and stabilities of
them are similar with these three. In the viewpoint of geometry,
the polarization states with a magnitude of 2P have a same space
group of Pm, and these polarization states with the magnitude of P
and $\sqrt{3}$P have a same space group of C2. Without changing the
relative position of any atom, through appropriate rotation and
reflection operation of $\vec{UP_1}+\vec{LP_1}$,
$\vec{UP_2}+\vec{LP_1}$ or $\vec{UP'_2}+\vec{LP_1}$, the other
twenty-seven polarization states can be achieved. For example, we
can rotate the whole structure of $\vec{UP_1}+\vec{LP_1}$ clockwise
by 120~degree to get $\vec{UP_2}+\vec{LP_2}$; do a reflection to
$\vec{UP_1}+\vec{LP_2}$ through a mirror plane of the boron atoms,
we can get $\vec{UP_2}+\vec{LP_1}$. Since there is no relative
space change of any atom during these rigid transformation, we
expect the electronic properties and stabilities will be kept after
these symmetry operation, and we do not show the band structures of
each one here.}

\section{Summary and Conclusions}

We study various properties of previously unexplored $\alpha$-phase
of 2D $MBOH$. In contrast to traditional FE materials with only
two polarization states, we identify eighteen stable polarization
states in this material. We propose a plausible scenario for
transformation between these polarization states, which is
corroborated by the vibration modes observed in the phonon
spectrum. Remarkably, the electric polarization state, which can be
tuned through controlling the $MOM$ chains, is found coupled with
the spin polarization. Furthermore, during the transformation of
the electrical polarization, ferroelasticity is observed.

\begin{table}[t]
\caption{%
\modR{The relative cohesive energy E for the relaxed structure with different
magnetic configurations in eV/unit-cell units. E$_1$ is the cohesive energy for the
structure optimized from the precursor MnB attached with  $-$OH radicals with
different magnetic configurations. E$_2$ is the relative cohesive energy for the
relaxed structure of the MBOH with different magnetic configurations.
The up and down arrows represent the two opposite spin states. E of the most stable structure is set to be 0.}%
}%
\begin{tabular}{lccc} %
\hline \hline
   \textrm{} %
 & \textrm{} %
 & \textrm{} %
 & \textrm{Configuration} %
 \\
  {Allotrope$\quad$} %
& {$E_{1}$$\quad$} %
& {$E_{2}$$\quad$} %
& {$Mn_{11}$, $Mn_{12}$, $Mn_{13}$, $Mn_{14}$} %
\\
  {} %
& {(eV)$\quad$} %
& {(eV)$\quad$} %
& {$Mn_{21}$, $Mn_{22}$, $Mn_{23}$, $Mn_{24}$} %
\\
\hline%
  {FM} %
& {$2.63$} %
& {$1.01$} %
& {$\uparrow$ $\qquad$$\uparrow$ $\qquad$$\uparrow$ $\qquad$$\uparrow$} %
\\
  {} %
& {} %
& {} %
& {$\uparrow$ $\qquad$$\uparrow$ $\qquad$$\uparrow$ $\qquad$$\uparrow$} %
\\
\hline%
  {A-AFM} %
& {$0.00$} %
& {$0.00$} %
& {$\uparrow$ $\qquad$$\uparrow$ $\qquad$$\uparrow$ $\qquad$$\uparrow$} %
\\
  {} %
& {} %
& {} %
& {$\downarrow$ $\qquad$$\downarrow$ $\qquad$$\downarrow$ $\qquad$$\downarrow$} %
\\
\hline%
  {F-AFM1} %
& {$2.78$} %
& {$0.52$} %
& {$\uparrow$ $\qquad$$\uparrow$ $\qquad$$\downarrow$ $\qquad$$\downarrow$} %
\\
  {} %
& {} %
& {} %
& {$\uparrow$ $\qquad$$\uparrow$ $\qquad$$\downarrow$ $\qquad$$\downarrow$} %
\\
\hline%
  {F-AFM2} %
& {$2.23$} %
& {$0.71$} %
& {$\uparrow$ $\qquad$$\downarrow$ $\qquad$$\uparrow$ $\qquad$$\downarrow$} %
\\
  {} %
& {} %
& {} %
& {$\uparrow$ $\qquad$$\downarrow$ $\qquad$$\uparrow$ $\qquad$$\downarrow$} %
\\
\hline%
  {F-AFM3} %
& {$1.63$} %
& {$0.71$} %
& {$\uparrow$ $\qquad$$\downarrow$ $\qquad$$\downarrow$ $\qquad$$\uparrow$} %
\\
  {} %
& {} %
& {} %
& {$\uparrow$ $\qquad$$\downarrow$ $\qquad$$\downarrow$ $\qquad$$\uparrow$} %
\\
\hline%
  {A-AFM1} %
& {$2.53$} %
& {$0.14$} %
& {$\uparrow$ $\qquad$$\uparrow$ $\qquad$$\downarrow$ $\qquad$$\downarrow$} %
\\
  {} %
& {} %
& {} %
& {$\downarrow$ $\qquad$$\downarrow$ $\qquad$$\uparrow$ $\qquad$$\uparrow$} %
\\
\hline%
  {A-AFM2} %
& {$2.23$} %
& {$0.49$} %
& {$\uparrow$ $\qquad$$\downarrow$ $\qquad$$\uparrow$ $\qquad$$\downarrow$} %
\\
  {} %
& {} %
& {} %
& {$\downarrow$ $\qquad$$\uparrow$ $\qquad$$\downarrow$ $\qquad$$\uparrow$} %
\\
\hline%
  {A-AFM3} %
& {$1.53$} %
& {$0.49$} %
& {$\uparrow$ $\qquad$$\downarrow$ $\qquad$$\downarrow$ $\qquad$$\uparrow$} %
\\
  {} %
& {} %
& {} %
& {$\downarrow$ $\qquad$$\uparrow$ $\qquad$$\uparrow$ $\qquad$$\downarrow$} %
\\
\hline \hline %
\end{tabular}
\label{table1}
\end{table}

\section*{Appendix}

\subsection{\modR{Strategy of exploring allotropes of magnetic MnBOH}}

\modR{Concerning the atomic structure is strongly coupled with the
magnetic configuration, start from the precursor MnB shown in
Fig.~\ref{fig8}(a), we optimize the atomic structure with eight
possible magnetic orders. Considering the periodicity of the
magnetic order, the atomic structure is optimized using a
2$\times$2 supercell which is four times bigger than the primitive
unit cell in Fig.~\ref{fig1}(a). There are four upper Mn atoms
(labeled as Mn$_{11}$, Mn$_{12}$, Mn$_{13}$ and Mn$_{14}$) and four
lower Mn atoms (labeled as Mn$_{21}$, Mn$_{22}$, Mn$_{23}$ and
Mn$_{24}$) are contained in the supercell. With respect to the most
stable one, the relative cohesive energy E$_1$ and magnetic
configuration of these allotropes are listed in Table I. We find
that the magnetic configuration of the most stable one is the so
called A-AFM with intralayer FM coupling and interlayer AFM
coupling, and we name this allotrope as $\alpha$-phase in the main
text.}

\modR{Since we first start from the precursor MnB, and with
different magnetic configuration, we optimize the structure and get
the most stable $\alpha$-phase. In case the allotropes with these
magnetic configurations are metastable resulted by the chosen
precursor of MnB decorating with $-$OH radicals, we also use the
optimized atomic structure of $\alpha$-phase MnBOH as a precursor
but with different magnetic configurations to check the relative
stability of the $\alpha$-phase MnBOH with the A-AFM magnetic order.
Same with the above strategy, we use a 2$\times$2 supercell
$\alpha$-phase MnBOH as shown in Fig.~\ref{fig8}(b) to do the
energy calculation and structure optimization with considering
other seven magnetic orders. The relative cohesive energy E$_2$ is
also listed in Table I. We see no matter which precursor we use,
the most stable $-OH$ decorating MnB is the $\alpha$-phase MnBOH.}

\subsection{Spin-polarized projected density of state}

In order to acquire the magnetic information, we project the
density of states onto each element as shown in Fig.~\ref{fig9}.
For B$_{1}$ and B$_{2}$ atoms, electrons are equally populated in
both channels of spin, which means that the B atoms are not
magnetic. For Mn$_{1}$, O$_{1}$ and H$_{1}$, majority electrons
occupy the spin-down channel, while for Mn$_{2}$, O$_{2}$ and
H$_{2}$, majority electrons occupy the spin-up channel. The DFT
calculation shows that the magnetic moment of Mn$_{1}$, O$_{1}$,
H$_{1}$ is 3.96~${\mu}_B$, 0.06~${\mu}_B$, 0.01~${\mu}_B$, while
that of Mn$_{2}$, O$_{2}$, H$_{2}$ is -3.96~${\mu}_B$,
-0.06~${\mu}_B$, -0.01~${\mu}_B$, respectively. So, for the system,
magnetic moments are mainly coming form the Mn atoms and arranged
antiferromagnetically between the upper and lower side of $nh-B$.

\begin{figure}[t]
\includegraphics[width=0.9\columnwidth]{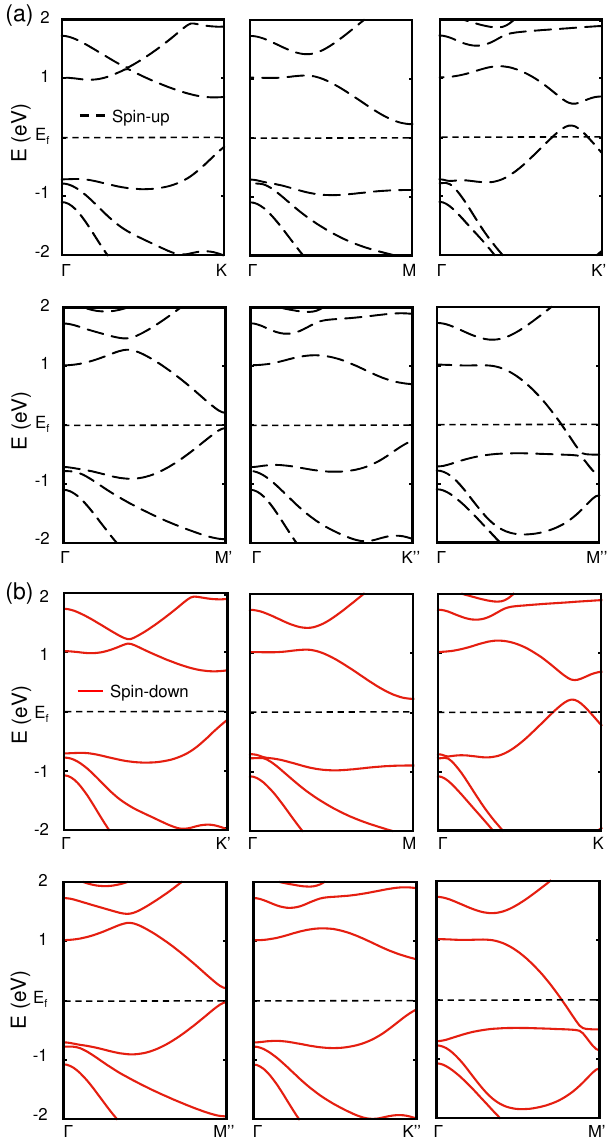}
\caption{%
Electronic energy-momentum dispersion relation from $\Gamma$ to each
$k$ point of $\alpha$-phase $MBOH$ with
(a) spin-up state,
(b) spin-down state.
} %
\label{fig11}
\end{figure}

\begin{figure}[t]
\includegraphics[width=0.7\columnwidth]{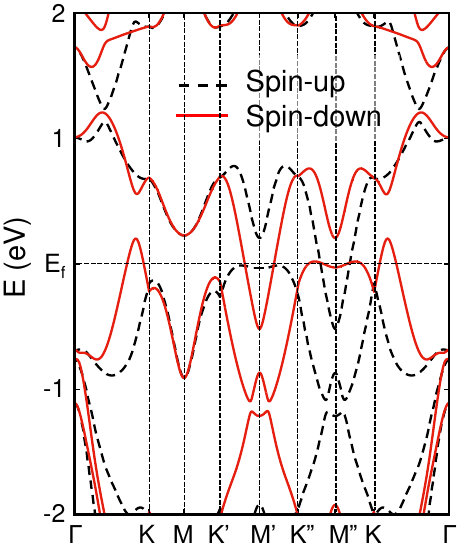}
\caption{%
Spin polarized electronic band structure of $\alpha$-phase %
$MBOH$ with electrical polarization status of $\vec{UP'_2}+\vec{LP_1}$.%
} %
\label{fig12}
\end{figure}

\begin{figure}[t]
\includegraphics[width=1.0\columnwidth]{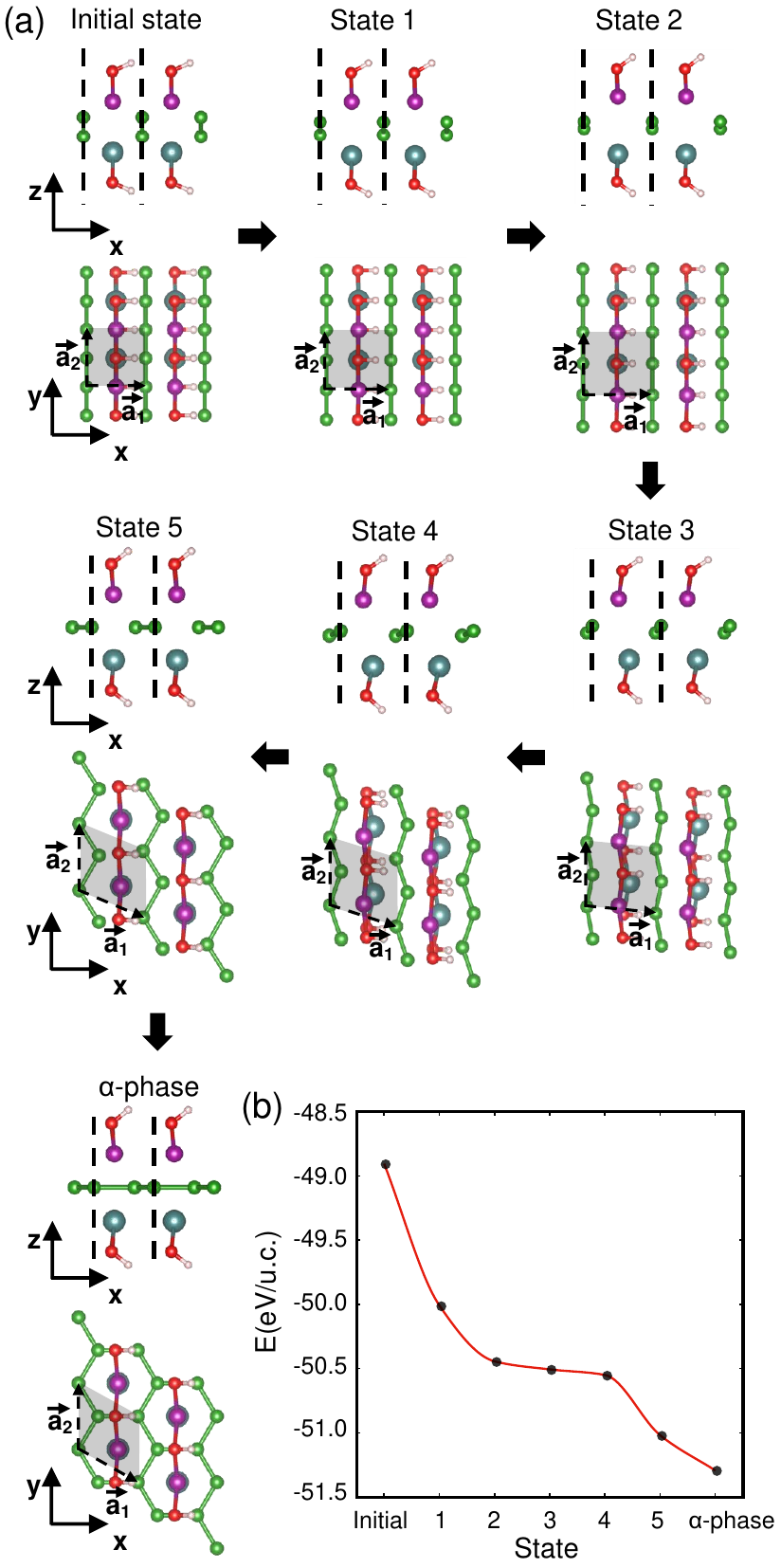}
\caption{(a) Schematic illustration of the atomic structure evolution from
the structure with $-OH$ radicals attached on the precursor MnB
to the stable
$\alpha$-phase $MBOH$ during the CG optimization. (b) Potential
energy E as a function of several picked states of $MBOH$ presented in
(a) during the structure optimization. %
\label{fig13}}
\end{figure}

\subsection{\modR{Thermal stability of polarization state of $\vec{UP_1}+\vec{LP_1}$,
            and dynamic stability of polarization states of
            $\vec{UP_2}+\vec{LP_1}$ and $\vec{UP'_2}+\vec{LP_1}$}}

\modR{As mentioned above, the $\alpha$-phase $MBOH$ can have thirty
polarization states, some of them have the same polarization
magnitude and direction but different atomic structures. Among
them, the $\vec{UP_n}+\vec{LP_n}$/$\vec{UP'_n}+\vec{LP'_n}$ (n=1,
2, 3) is the most stable one and the others are less stable by
75~meV/u.c.. In the main manuscript, we show the dynamic stability
of $\vec{UP_n}+\vec{LP_n}$, here, we have performed canonical
molecular dynamics (MD) simulation of the $\vec{UP_1}+\vec{LP_1}$
at room temperature (T=300 K) for a time period of 10~ps. As seen
in Figure~\ref{fig10}(a), the potential energy fluctuates around a
constant value, indicating that the structure are dynamically
stable at room temperature. As shown in Figure~\ref{fig10}(b), we
have also calculated phonon spectra of $\vec{UP_2}+\vec{LP_1}$ and
$\vec{UP'_2}+\vec{LP_1}$ as representative of these less stable
ones. All spectra are free of imaginary frequencies, meaning that
expect the most stable ones of
$\vec{UP_n}+\vec{LP_n}$/$\vec{UP'_n}+\vec{LP'_n}$, the other
energetic less stable polarization states are also are dynamically
stable.}

\subsection{Decomposition of Electronic Band Structure of $\alpha$-phase $MBOH$ with Polarization State $\vec{UP_2}+\vec{LP_1}$}

As we discussed, the splitting of band structure for $\alpha$-phase
$MBOH$ with polarization status of $\vec{UP_2}+\vec{LP_1}$ is
coming from the transformation between corresponding reciprocal
lattices of the sub atomic structures with spin-up and spin-down
states. The electronic properties for the sub atomic structure with
spin-up at momentum of $K$, $K'$, $M'$, $M"$ are equivalent with
that for the sub atomic structure with spin-down at momentum of
$K'$, $K$, $M"$, $M'$. We show the electronic energy-momentum
dispersion relations of spin-up and spin-down states from $\Gamma$
to each $k$ point in Figures~\ref{fig11}. We see the dispersion
relation of spin-up state in the path of $\Gamma$-$K$,
$\Gamma$-$K'$, $\Gamma$-$M'$, $\Gamma$-$M"$ in
Figures~\ref{fig11}(a) are the same with dispersion relation of
spin-down state in the path of $\Gamma$-$K'$, $\Gamma$-$K$,
$\Gamma$-$M"$, $\Gamma$-$M'$ in Figures~\ref{fig11}(b)
correspondingly, which proves the validity of our conclusion.

\subsection{Electronic Band Structure of $\alpha$-phase $MBOH$ with Polarization of $\vec{UP'_2}+\vec{LP_1}$}

We calculate the electronic band structure of $\alpha$-phase $MBOH$
with polarization status of $\vec{UP'_2}+\vec{LP_1}$ as shown in
Figure~\ref{fig12}. The difference in atomic structure of
$\alpha$-phase $MBOH$ between the polarization status of
$\vec{UP'_2}+\vec{LP_1}$ and $\vec{UP_2}+\vec{LP_1}$ is the
orientation of $-OH$ radicals, but the spin polarized band
structure of the two polarization status are identical. In other
words, as we mention in the main manuscript, whether the H atoms
are staying at the right side or left side of the $MOM$ chain has
no effect on the band structure.

\subsection{Formation of $\alpha$-phase $MBOH$}

In main manuscript, we mentioned that a 2D boron with
nearly-honycomb lattice is formed in $\alpha$-phase $MBOH$, which
is meaningful for studying the superconductivity in boron system.
For real experiment, we need to analysis the feasibility of
synthesising such material. In our calculation, we use the CG
method to do the structure optimization. This method can provide a
figure of merit about the evolution from the initial state to the
finial optimized state. We present several snap shots of the system
during the optimization process in Figure~\ref{fig13}(a) to
illustrate the formation of the $\alpha$-phase $MBOH$ from the
precursor MnB with $-OH$ radicals attached. Corresponding with
these states, the potential energy $E$ of the system is shown in
Figure~\ref{fig13}(b). We see that from the initial state to the
$\alpha$-phase, $E$ decreases monotonically, which indicates
$\alpha$-phase is relatively easy to be synthesised in experiment
using $-OH$ radicals to decorate the MnB.

\bigskip
\begin{acknowledgements}
This study is supported by the Natural Science Foundation of the
Jiangsu Province (Grant No.\ BK20210198), the National Natural
Science Foundation of China (Grant Nos. 62274028, 12204095 and
12104090), the High Level Personnel Project of Jiangsu Provience
Grant No.\ JSSCBS20220120, `Zhishan' Scholars Program of Southeast
University (Grant Nos. 2242023R10006 and 2242021R40004), open
research fund of Key Laboratory of Quantum Materials and Devices
(Southeast University) of Ministry of Education, and the
Fundamental Research Funds for the Central Universities (Southeast
University).

Pingwei Liu and Dan Liu contributed equally to this work.

\end{acknowledgements}

%

\end{document}